\documentstyle[preprint,aps]{revtex}
\input epsf.tex
\def\DESepsf(#1 width #2){\epsfxsize=#2 \epsfbox{#1}}
\def\joinp{\mathrel{\mkern-14mu}}
\def\slashR{\mathrel {\rm R} \joinp \not \ \ }
\begin{document}
\preprint{\vbox{\hbox{}}}
\title{Neutrino mass induced radiatively by supersymmetric leptoquarks}
\author{Chun-Khiang Chua\thanks{Email address: ckchua@phys.ntu.edu.tw},
        Xiao-Gang He, and W-Y. P. Hwang}
\address{
Department of Physics, National Taiwan University,\\
Taipei, Taiwan 106, Republic of China\\
}

\maketitle
\begin{abstract}
We show how nonzero Majorana neutrino masses 
can be radiatively generated
by extending the MSSM with leptoquark chiral multiplets 
without violating R-parity.
It is found that, with these particles,
the R-parity conservation does not imply lepton number conservation. 
Neutrino masses generated at a one-loop level
are closely related to the down quark mass matrix.                       
The ratio of neutrino mass-squared splittings 
$\Delta m^2_{\nu_2-\nu_1}/\Delta m^2_{\nu_3-\nu_2}$ obtained 
is naturally close to $\Delta m^2_{s-d}/\Delta m^2_{b-s}\sim 10^{-3}$
which is in the right region required to explain both the
atmospheric neutrino data and the MSW solutions for the solar neutrino data.

\end{abstract}


\preprint{\vbox{\hbox{}}}

\preprint{\vbox{\hbox{}}}


\section{Introduction}

Many recent experimental observations have hinted at 
non-vanishing neutrino masses\cite{11c}.
The strongest hint is from the Super-Kamiokande underground experiment
\cite{SuperK}.
They observed an up-down asymmetry of atmospheric $\nu_\mu$ neutrino flux,
suggesting an oscillation of $\nu_\mu$ into $\nu_\tau$ (or $\nu_s$).
The mass squared splitting is found in the range of
 $1.5\times 10^{-3}\leq\Delta m^2_{\nu_3-
\nu_2}\leq 6\times 10^{-3} {\rm eV}^2$ (with 95\% C.L.),
and the mixing angle is close to the maximal mixing\cite{fisher}. 

In addition, 
the observed solar neutrino fluxes \cite{sol} 
are well below the Standard Solar Model predictions.
The details may be explained by invoking neutrino mixing between $\nu_e$ and
$\nu_\mu$.
There are two types of solutions. 
One is the solution due to
the MSW mechanism\cite{MSW}
with either large or small mixing angles\cite{bestfit}.
For the small mixing angle (SMA) solution, one needs
$\Delta m_{\nu_2-\nu_1}^2=(0.4-1)\times10^{-5} {\rm eV}^2$, and 
$\sin^2 2\theta=(0.12-1.2)\times10^{-2}$. 
For the large mixing angle (LMA) solution,
the mass splitting and mixing angle are
$\Delta m^2_{\nu_2-\nu_1} = (0.6-20)\times 10^{-5} {\rm eV}^2$, and
$\sin^2 2\theta \sim 0.76$.
The other one is the so-called just-so solution, 
which corresponds to a neutrino oscillation in vacuum \cite{justso}.
For this case, one needs 
the mass-squared splitting $\Delta m^2=8.0\times 10^{-11} {\rm eV}^2$ and    
the mixing angle $\sin^2 2\theta=0.75$ 
for the best fit solution \cite{bestfit}.

If neutrinos have masses, they are expected to be small
from laboratory bounds and cosmological consideration 
\cite{pdg,nubb,cosbound}.
In the Standard Model (SM) neutrinos are massless.
One needs to
extend the SM to accommodate massive neutrinos. 
A neutrino mass can be generated by a Higgs Yukawa interaction 
like any other fermions if 
right-handed neutrinos are introduced.
However, this corresponds to an unnaturally small Higgs Yukawa coupling,
$\lambda\sim m_{\nu}/v \leq {\rm eV}/{246\,{\rm GeV}}\sim 10^{-11}$,
where $v$ is the vacuum expectation value of the Higgs field.
There have been extensive studies of neutrino masses in models beyond the SM 
\cite{moha,mixing}.
To explain the smallness of the neutrino mass,
a popular approach is to 
use the see-saw mechanism to trade a light neutrino with
a super heavy neutrino \cite{seesaw} or 
to radiatively generate neutrino masses such that the 
smallness of neutrino masses are due to the loop suppression \cite{radmass}.

In studying neutrino masses, it is interesting to note that
the ratio of the mass-squared splittings 
$\Delta m_{\nu_2-\nu_1}^2/\Delta m_{\nu_3-\nu_2}^2$ 
for the MSW solutions to the solar 
neutrino problem can be
close to $\Delta m_{s-d}^2/\Delta m_{b-s}^2\approx 10^{-3}$.
This may be regarded as a hint 
that the neutrino mass matrix is proportional to
the down quark mass matrix. In this case the heaviest neutrino has 
a mass less than $\sim 0.08$ eV. It is interesting to see if 
such a mass matrix can be obtained in a natural way. 
In this paper we study a model in which the Majorana neutrino
mass is radiatively 
generated by supersymmetric leptoquarks and the mass matrix
is closely related to the down quark mass matrix. 

The supersymmetry (SUSY) is one of the leading candidates for physics beyond 
the SM\cite{susypdg}. 
In order to generate neutrino masses one needs to violate R-parity
and/or introduce new physics \cite{R-parity,ema}. 
We will show that it is possible to generate Majorana neutrino masses 
without the violation of R-parity if leptoquarks are introduced in the model.
In the presence of these leptoquarks,
the lepton number ($L$) can be broken explicitly 
even if the R-parity is conserved.
The lepton number violating interaction can generate Majorana neutrino masses
at a one-loop level and naturally lead to a mass matrix proportional to
the down quark mass matrix.
This is due to a special feature of the scalar leptoquark interaction.
It can be shown that a lepton number violating vector leptoquark model
will relate a neutrino mass matrix to the up quark mass matrix,
which would not naturally give neutrino mass hierarchies mentioned above. 
In the SUSY extension with the SM gauge group, 
vector leptoquarks are not allowed. This provides
additional motivation to study SUSY models with leptoquark multiplets.

\section{The model} 

The matter and Higgs superfields in the MSSM are
$\hat L$, $\hat E$, $\hat Q$, $\hat U$, $\hat D$, $\hat H_1$ and $\hat H_2$.
Their quantum numbers under SU(3$)_C\times$SU(2$)_L\times$U(1$)_Y$ are
(1, 2, -1/2), (1, 1, +1), (3, 2, +1/6), ($3^*$, 1, -2/3), ($3^*$, 1, +1/3),
(1, 2, -1/2) and (1, 2, +1/2), respectively.
To avoid lepton number and baryon number violations,
R-parity is introduced.
The ordinary fields, the quarks, leptons and Higgses, are assigned with 
positive R-parity and their super-partners are assigned with negative 
R-parity. There are in general seven types of leptoquarks 
which can couple directly to the quark and lepton field 
products, $\hat U\times \hat E$, $\hat Q\times \hat L$, $\hat D\times \hat E$,
$\hat U\times \hat L$, $\hat Q\times \hat E$, and $\hat D\times \hat L$, 
at the tree level 
with their gauge quantum numbers
given, respectively, by \cite{buch}
\begin{eqnarray}
\label{2}
&\quad\hat S:\,(3,\, 1,\,-1/3),\,\,                                          
\hat S':\,(3^*,\, 1,\, +1/3),\,\,
\hat T:\,(3^*,\, 3,\, +1/3),\,\,
\hat S^{''}:\,(3,\, 1,\, -4/3),\,\,
\nonumber \\
&\hat R:\, (3,\, 2,\, +7/6),\,\,
\hat R':\,(3^*,\, 2,\, -7/6),\,\,
\hat I:\,(3,\, 2,\, +1/6).                                        
\end{eqnarray}
The R-parity assignments of these leptoquark supermultiplets are determined
by their interaction with lepton and quark multiplets to be the same as that
of Higgs multiplets.

For the purpose of generating Majorana neutrino masses 
only two types of leptoquarks are needed which are
$\hat S'$ and $\hat I$. 
These leptoquarks provide two units of the lepton number violation
needed for  Majorana neutrino masses through the R-parity conserving
interaction in the superpotential by a term,
$\epsilon_{ij} \hat H^i_1 \hat I^j \hat S'$.
In order to cancel the 
gauge anomalies due to 
the fermion partners of leptoquarks for consistency
more leptoquark multiplets are required.
We note that the quantum number of $\hat S$ is conjugate to that of 
$\hat S^\prime$ and accordingly
it can be used to cancel the anomalies due to
$\hat S'$.
We need to introduce a new leptoquark multiplet $\hat I^\prime$
with quantum number conjugates to $\hat I$ to cancel the anomalies.
We also note that the R-parity does not forbid a baryon number violating 
term $\hat U\hat D\hat S^\prime$. To avoid rapid proton decay, we must
either assume the coupling is very small, or introduce further
symmetry. 
To make use of the symmetry, 
we may assign a negative $Z_2$ parity to 
$\hat U$, $\hat D$ and $\hat Q$, while all other matter fields in the MSSM
have a positive $Z_2$ parity. 
The $Z_2$ parity of leptoquarks is negative as determined
from leptoquark interactions.
Thus, the $Z_2$ symmetry will forbid the baryon number violating terms.
In short, we are considering the leptoquark interaction instead of
the diquark interaction.

\vspace{0.2 in}
\begin{tabular}{llll}                                                          
Superfield$\qquad$                                                             
                &SU(3)$\times$SU(2)$\times$U(1)$\times L$$\qquad$         
                                     &Boson Fields$\qquad$                  
                                                    &Fermionic Partners     
\\ \hline                                                                      
$\hat S$        & (3, 1, -1/3; -1)    &$ S_L$        &$\tilde S_L$            
\\                                                                             
$\hat S^\prime$ &($3^*$, 1, +1/3; +1)  &$S^*_R$       &$\tilde S^c_L$         
\\                                                                        
$\hat I$
                & (3, 2, +1/6; +1)   &$(O_L,P_L)$   &$(\tilde O_L,\tilde P_L)$ 
\\
$\hat I^\prime$        &($3^*$, 2, -1/6; -1)     &$(P^*_R,-O^*_R)$           
                                             &$(\tilde P^c_L,-\tilde O^c_L)$
\end{tabular}

\vspace{0.1 in}
\noindent 
\vspace{-0.1 in} {\bf Table 1}. The particle content of             
leptoquark chiral supermultiplets related to the neutrino mass generation.
The $L$ denotes the lepton number.
\vspace{0.2 in}

The model we now have is an extended MSSM 
with the standard particle content plus leptoquarks and leptoquarkinos in 
Table 1. 
With these multiplets we can construct the 
leptoquark R-parity conserving superpotential $W_{LQ}$ with  
\begin{eqnarray}
W_{LQ} &=&-(h^*)_{ab}\, \hat U_a \hat E_b \hat S
  -(h^\prime)_{ab}\, \epsilon_{ij} \hat Q^i_a \hat L^j_b \hat S^\prime
  -(h^{\prime\prime})_{ab}\,\epsilon_{ij} \hat D_a \hat L^i_b \hat I^j
\nonumber \\
&&+g\, \epsilon_{ij} \hat H^i_1 \hat I^j \hat S^{\prime}
  +M_S \hat S_d \hat S^\prime_d+ M_I \epsilon_{ij} 
                                      \hat I^{\prime i}\hat I^j, 
\label{WLQ}
\end{eqnarray}
where $a$ and $b$ are generation indices of quarks and leptons in 
the weak basis.
Note that $\tilde S_R$ and $\tilde P_L$ can be mixed by the $g$ term
when the $H_1$ takes its vacuum expectation value (VEV) $v_1$;
otherwise,  the masses are $M_S$ and $M_I$ respectively.

With this superpotential and the Higgs VEVs, we get leptoquark mixings of
$S_L$ with $P_L$, $S_R$ with $P_R$, and $S_R$ with $P_L$. 
The F-term causes mixings, 
and the D-term contributes to diagonal parts of the mass matrix only.
The mixing terms are
\begin{equation}
V=gv_1\,(M_S\,S_L^\dagger P_L+M_I\,S_R^\dagger P_R
    -\mu\tan\beta\,S_R^\dagger P_L)+h.c.
\end{equation}

There are more terms due to the SUSY soft breaking, 
\begin{eqnarray}
\label{softmass}
V_{\rm soft}&=&{1\over 2} M^2_{S_R}\, S^\dagger_R S_R
    +{1\over 2} M^2_{S_L}\, S^\dagger_L S_L
    +{1\over 2} M^2_{I_L}\, I^\dagger I
    +{1\over 2} M^2_{I_R}\, I^{\prime\dagger} I^\prime
\nonumber \\
  &&+M^2_{S_R-S_L}\,S^\dagger_R S_L
    +M^2_{I-I^\prime}\,\epsilon_{ij} I^iI^{\prime j}
    +M_{HIS} \epsilon_{ij} H^i_1 I^j S^\dagger_R 
    +h.\,c.
\end{eqnarray}
Note that the (standard) SUSY soft breaking terms do not include
$H_1 I S^\dagger_L$ or $H^\dagger_2 I S^\dagger_R$ and so on \cite{soft}.
The mass parameters $M_i$ in the Eq. (\ref{softmass}) 
are naturally in the SUSY breaking scale. 
The $M_{HIS} \epsilon_{ij} H^i_1 I^j S^\dagger_R$ term 
violates lepton number by two units. After spontaneous symmetry breaking this 
term produces the mixing of $S_R$ and $P_L$ which plays a crucial role in 
generating Majorana masses at a one-loop level. This is the dominant source for
the lepton number violation related to the Majorana neutrino masses although
other leptoquarks contribute through mixings among 
$S_L$, $S_R$, $P_L$ and $P_R$. 
We note that the mass parameters, $M_S$ and $M_I$,
from the superpotential are free parameters which may in general 
be smaller or larger than the SUSY breaking scale $M_{SUSY}$,
while those mass parameters contained in soft breaking terms are
naturally in the scale of $M_{SUSY}$.

\section{Radiative Majorana neutrino masses}

The Majorana neutrino mass term 
${1/ 2}\,({\bar \nu^c_{Ra}}\,m_{ab}\,{\nu_{L_b}}+h.c.)$
can be obtained from one-loop diagrams with all possible leptoquark 
mass insertions in
the internal leptoquark line and the internal quark line.
The leading contribution comes from a  mass insertion each in the quark line
and the other insertion in the leptoquark line as shown in Figure 1.
The mass matrix $m_{ab}$ is symmetric and the leading term is given by,
\begin{equation}
m_{ab} = {N_c\over{ 8\pi^2}} 
  {\Delta {\cal M}^2 \over  ({\bar M_{LQ}}^2) }
 {1\over2}(h^{\prime T} M_D h^{\prime\prime}
                 +h^{\prime\prime T} M^T_D h^{\prime})_{ab},
\label{mass}
\end{equation}
where $N_c=3$ is the number of color, $M_D$ is the down type quark mass matrix,
${\bar M_{LQ}}$ is the average leptoquark mass and 
$\Delta {\cal M}^2={\cal M}^2_{P_L-S_R}
   =-gv_1\mu\tan\beta+M_{HIS}\,v_1
   \sim M_{SUSY}\, v$
denotes the $P_L-S_R$ mixing element as shown in Eq. (3) and Eq. (4).

The quark mass matrix can be diagonalized in the usual way,
$V_{dL} M_D V^\dagger_{dR}=M^{\rm diag}_D$,
where the diagonal mass matrix is given by\cite{gasser},
\begin{eqnarray}
M^{\rm diag}_D(\mu)&=&{\rm diag}(m_d,\,m_s,\,m_b)
\nonumber \\
&=&m_b(\mu)\, {\rm diag}\Bigl(
(1.36\pm0.42)\times10^{-3},\,(2.66\pm0.87)\times 10^{-2},\,1\pm0.05
\Bigr)\,.
\end{eqnarray}
From Eq. (\ref{mass}), 
it is clear that the neutrino mass matrix obtained is closely related
to the down quark mass matrix.
We will work in a quark field basis where the down quark mass matrix is
hermitian, which can always be done\cite{jars}.
If small CP violation entries are neglected the down quark mass matrix is
symmetric, $V_{dL}$ becomes orthogonal and equals to $ V_{dR}$.
This choice of basis is convenient for our purpose
since the Majorana neutrino mass matrix is symmetric.

To finally obtain numerical numbers for neutrino masses,
we need to know parameters related to leptoquarks.
The couplings $h^\prime,\;h^{\prime\prime}$ and the leptoquark mass $M_{LQ}$ 
relevant to neutrino masses in our model are
unknown which must be subjected to experimental constraints. 
The typical experimental lower bound on the leptoquark mass is about
200 GeV \cite{direct}.  
There are many studies on leptoquark couplings and masses from indirect
searches \cite{pdg,indirect1,indirect2,constraint}.
We now consider the constraints on 
$h^\prime_{ab},\,h^{\prime\prime}_{ab}$ and $M_{LQ}$.
We follow ref. (\cite{indirect1}) to evaluate te constraints.
We note that $h^\prime,\,h^{\prime\prime}$ are denoted as
$\lambda_{LS_o},\,\lambda_{L\tilde S_{1/2}}$ in
ref. \cite{indirect1}.

For $h^\prime_{ab}$ the strongest constraint comes from 
the lepton flavor-changing rate for a muon to turn into an electron when 
scattered off a nucleus, such as $\mu{\rm Ti}\to e{\rm Ti}$. 
This gives the constraint on the coupling of $S_R$ as \cite{indirect1}
\begin{equation}
\label{Ti}
\sqrt{h^\prime_{11} h^\prime_{12}}/M_{LQ}
\leq 0.01\,{\rm TeV}^{-1}.
\end{equation}
Similar consideration gives a constraint on 
the coupling of $I^i$
with $\sqrt{h^{\prime\prime}_{11} h^{\prime\prime}_{12}}/M_{LQ}
\leq 0.01\,{\rm TeV}^{-1}$.
The leptoquark exchanges also contributes to
the kaon semileptonic decay, $K^+\to\pi^+\nu\bar\nu$. 
By requiring that the prediction is smaller then the observed bound of
$Br(K^+\to\pi^+\nu_l\bar\nu_{l^\prime})=0.42^{+0.97}_{-0.35}\times10^{-9}$
gives constraints \cite{indirect1,BNL}, 
\begin{equation}
\label{kpinubarnu}
\sqrt{h^{\prime}_{1l} h^{\prime}_{2l^\prime}}/M_{LQ},\,
\sqrt{h^{\prime\prime}_{1l} h^{\prime\prime}_{2l^\prime}}/M_{LQ}
\leq 0.03\,{\rm TeV}^{-1}.
\end{equation}
More stringent constraints on the $I^i$ couplings come from
lepton family number violating leptonic kaon decay, $K^0_L\to\mu^\pm e^\mp$.
Using the constraint in ref. \cite{indirect1} and the new experimental result
\cite{pdg},
$Br(K^0_L\to\mu^\pm e^\mp)\leq 3.3\times 10^{-11}$, 
we have
\begin{equation}
\label{kmue}
\sqrt{h^{\prime\prime}_{11} h^{\prime\prime}_{22}}/M_{LQ},\,
\sqrt{h^{\prime\prime}_{12} h^{\prime\prime}_{21}}/M_{LQ}
\leq 6\times10^{-3}\,{\rm TeV}^{-1}.
\end{equation}
One can find constraints on other elements of 
$h^\prime_{ab},\,h^{\prime\prime}_{ab}$ in ref. \cite{indirect1},
they are in general weaker than the ones discussed here.

As we turn on the mixing of $S_R$ with $P_L$, 
these chiral leptoquarks become non-chiral leptoquarks,
there are further constraints for them.
The constraints from the chirally suppressed pion decay,
$\pi^+\to e^+\nu_l$ give \cite{pdg,constraint},
\begin{equation}
\label{pienu}
\sqrt {h^\prime_{11} h^{\prime\prime}_{1l}} 
{\sqrt{M_{SUSY} v}\over M^2_{LQ}},\,
\sqrt {h^\prime_{1l} h^{\prime\prime}_{11}} {\sqrt{M_{SUSY} v}\over M^2_{LQ}}
\leq 4\times 10^{-3}\,{\rm TeV}^{-1}.
\end{equation}
The similar kaon leptonic decay gives a weaker constraint due to
$m_s >m_u,\,m_d$.
The mixing of $S_R$ with $P_L$ contributes to the lepton number violating
($\Delta L=2$) $K^+\to \pi^+\nu_l\nu_{l^\prime}$ decay. 
This gives
\begin{equation}
\label{kpinunu}
\sqrt{h^\prime_{1l} h^{\prime\prime}_{2l^\prime}} 
{\sqrt{M_{SUSY} v}\over M^2_{LQ}},\,
\sqrt{h^\prime_{2l} h^{\prime\prime}_{11^\prime}} 
{\sqrt{M_{SUSY} v}\over M^2_{LQ}}
\leq 2\times 10^{-3}\,{\rm TeV}^{-1}.
\end{equation} 
The factor $\sqrt{M_{SUSY} v}/M_{LQ}$ in Eqs. (\ref{pienu}) 
and (\ref{kpinunu})
reminds us that these non-chiral leptoquark behavior is due to the
mixing and the constraints are weaker than the constraints obtained in
Eq. (\ref{kmue}) if $\sqrt{M_{SUSY} v}/M_{LQ} <<1$.
Without any mixing, there is no non-chiral leptoquark in the SUSY model. 

To estimate what value of $m_{ab}$ may be,
we may work with a simplified situation for illustration, 
with $h^{\prime}_{ij}\sim h^{\prime\prime}_{ij}\sim h\, \delta_{ij}$.
We obtain a very simple and very interesting  mass matrix with
the mass eigenvalues of the
neutrinos given by
\begin{equation}
(m_{\nu_1},\,m_{\nu_2},\, m_{\nu_3}) = {N_c\over{ 8\pi^2}} 
  {\Delta {\cal M}^2 \over {\bar M_{LQ}}^2 }
  h^2 (m_d,\,m_s,\,m_b),
\end{equation}
and the mixing matrix is given by $V_{dL}$.
We note that the small mixing behavior of Cabbibo-Kobayashi-Maskawa 
matrix ($V_{CKM}=V_{uL}V^\dagger_{dL}$) does not
necessarily imply any small mixing 
pattern in the down quark mass matrix or up quark mass matrix separately.
There are parameter spaces in which the
required mixing angles for the atmospherical and the solar neutrino data 
can be produced.

To obtain $\Delta m^2_{\nu_3-\nu_2}$ to be below $6\times 10^{-3}$ eV$^2$
as required by the atmospheric neutrino data,
$h/M_{LQ}$ is restricted to be less than $0.55\times 10^{-4}\,{\rm TeV}^{-1}$ 
which is safely
below the above constraints for $M_{LQ}\geq M_{SUSY}$.
We note that for the particular case where $h^\prime$ and $h^{\prime\prime}$
are universal and flavor blind,  constraints in 
Eqs. (\ref{kpinubarnu}), (\ref{kmue}), (\ref{pienu}) and (\ref{kpinunu}) all
apply. 
Using $h/M_{LQ}=0.53\times10^{-4}\, {\rm TeV}^{-1}$ 
with $m_b$ at the top quark mass scale and Eq. (\ref{mass}),
we obtain Majorana neutrino masses,
\begin{eqnarray}
m_{\nu_1}&=&(9.92\pm3.08)\times10^{-5} {\rm eV},
\nonumber \\
m_{\nu_2}&=&(1.94\pm0.64)\times10^{-3} {\rm eV},
\nonumber \\
m_{\nu_3}&=&(7.30\pm0.36)\times10^{-2} {\rm eV}.
\end{eqnarray}
These values correspond to mass-squared splittings of
\begin{eqnarray}
\Delta m^2_{\nu_3-\nu_2}
&=&(5.32\pm0.53)\times10^{-3} {\rm eV}^2,
\nonumber \\
\Delta m^2_{\nu_2 -\nu_1}&=&
(3.75\pm2.48)\times10^{-6} {\rm eV}^2,
\end{eqnarray}
which are in the right regions with the atmospheric neutrino solution 
and the MSW solar neutrino small mixing angle and large mixing angle solutions.
These Majorana neutrino masses are well below the bound
on the effective Majorana mass ($m_{\nu_e\nu_e}\leq 0.2\, {\rm eV}$) 
from the negative result of the neutrinoless double beta decay \cite{nubb}.

If we use vector leptoquarks instead of scalar leptoquarks,
we would have obtained neutrino masses proportional to up type quark masses.
There is an important difference when compared with the scalar leptoquark case
discussed above.
One would naturally obtain $(m_{\nu_1},\,m_{\nu_2},
\,m_{\nu_3})$ to be proportional to
$(m_u,\,m_c,\,m_t)$. The ratio
$\Delta m^2_{\nu_2-\nu_1}/ \Delta m^2_{\nu_3-\nu_2}$ is then
given by $\Delta m^2_{u-c}/\Delta m^2_{c-t} \sim 2\times 10^{-5}$ which 
disagrees with the data.

\section {Conclusion}

In this paper, 
we have shown that a simple 
extension of the MSSM with leptoquark multiplets
contains a lepton number violating interaction which can generate 
Majorana neutrino masses
at a one-loop level without the breaking of R-parity.
One may contrast this scenario to the R-parity violating extension of MSSM.
In both cases, in order to obtain a light neutrino mass, one needs the ratio
$h/M$ (or $\lambda_{\slashR}/M_{MSSM}$ in the extended MSSM) 
to be of the order of $10^{-4}\,{\rm TeV}^{-1}$.
In the particle content of the MSSM all particle masses ($M_{MSSM}$)
should be below $M_{SUSY}$, and this would correspond to suppressed 
R-parity violating Yukawa couplings ($\lambda_{\slashR}$).
In our case, in the presence of the leptoquark multiplets,
the Yukawa couplings can in general be not artificially suppressed.
The $\Delta L=2$ interaction in this model is favored than the $\Delta L=1$
interaction in the extended MSSM in light of leptogenesis \cite{ema}.
Furthermore,
the radiatively generated neutrino mass matrix in this model is closely 
related to the down quark mass matrix as leptoquarks couple to leptons and 
down quarks. 
In the simplest scenario with the coupling matrices for
leptoquarks, quarks and leptons being proportional to unit matrix, that is, 
$h'_{ij}\sim h''_{ij}\sim h\,\delta_{ij}$, it predicts
a hierarchy structure in neutrino masses
$\Delta m^2_{\nu_2-\nu_1}/ \Delta m^2_{\nu_3-\nu_2}
\sim \Delta m^2_{d-s}/\Delta m^2_{s-b}$ which
provides mass-squared splittings required to explain both
the atmospheric neutrino and the MSW solar neutrino data.

More complicated structures for $h'$ and $h''$ can lead to different
mass hierarchy patterns. 
The simple neutrino mass matrix that is proportional 
to the down quark mass matrix, 
depends on that the matrices $h'(h'')$ being proportional to 
the unit matrix. 
It remains to be seen that if 
this can be obtained from some symmetry principle.

This work
was supported in part by a grant from National Science Council of Republic of
China grants NSC88-2112-M002-001Y and NSC88-2112-M002-041.
%
%

\begin{figure}[htb]                                                             
\centerline{\DESepsf(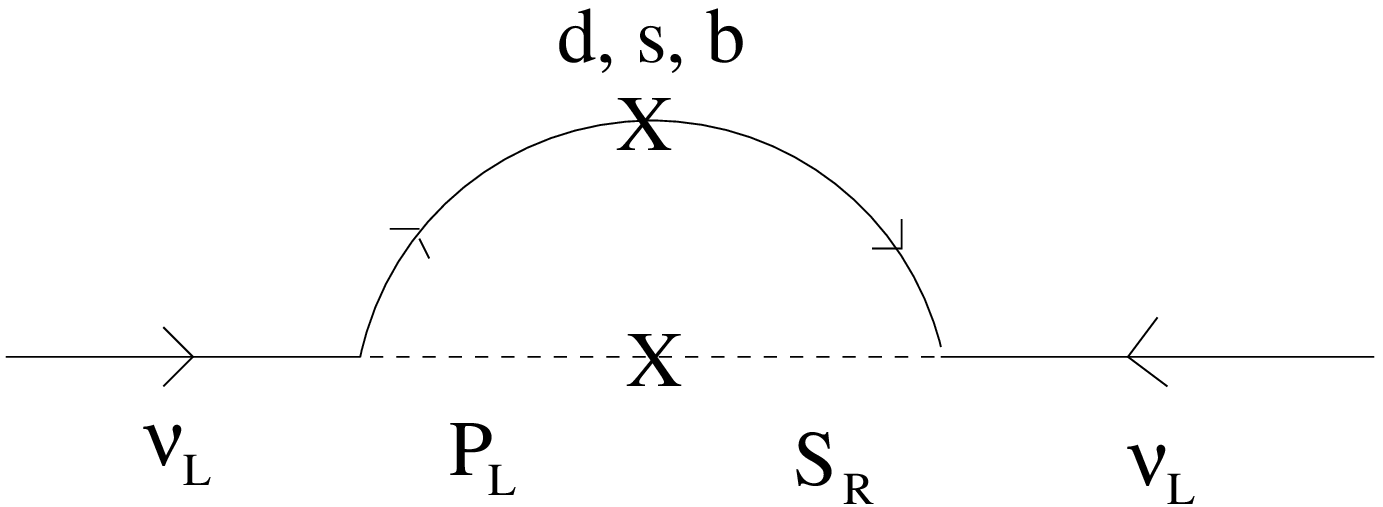 width 12cm)}                                        
\smallskip                                                                      
\caption{The leading one-loop diagram which gives rise to 
nonzero neutrino masses, the {\bf X} represent a mass  insertion.} 
\end{figure}                                                                    

\end{document}